%% file: Manuscript.tex
\documentclass[preprint, 12 pt]{elsarticle}

\usepackage{graphicx}
\usepackage{hyperref}
\usepackage[utf8]{inputenc}
\usepackage{multirow}
\usepackage{array,multirow}
\usepackage{hyperref}
\usepackage{caption}
\usepackage{subcaption}
\usepackage[legalpaper, margin=1in]{geometry}

\usepackage{amssymb}
\usepackage{amsmath}
\usepackage{subcaption}

\journal{arXiv}
\begin{document}

\begin{frontmatter}


\title{A Deep Learning Study on Osteosarcoma Detection from Histological Images}



\author[address1]{D M Anisuzzaman}
\ead{anisuzz2@uwm.edu}

 \author[address1]{Hosein Barzekar\corref{corresponding author}}
\cortext[corresponding author]{Corresponding author}
\ead{barzekar@uwm.edu}

\author[address2]{Ling Tong}
\ead{ltong@uwm.edu}

 \author[address2]{Jake Luo}
\ead{jakeluo@uwm.edu}

\author[address1,address3]{Zeyun Yu}
\ead{yuz@uwm.edu}

\address[address1]{Big Data Analytics and Visualization Laboratory, Department of Computer Science, University of Wisconsin-Milwaukee, Milwaukee, WI 53211, USA}
\address[address2]{Department of Health Informatics and Administration, University of Wisconsin-Milwaukee, Milwaukee, WI 53211, USA}
\address[address3]{Department of Biomedical Engineering, University of Wisconsin-Milwaukee, Milwaukee, WI 53211, USA}

\begin{abstract}
In the U.S, 5-10\% of new pediatric cases of cancer are primary bone tumors. The most common type of primary malignant bone tumor is osteosarcoma. The intention of the present work is to improve the detection and diagnosis of osteosarcoma using computer-aided detection (CAD) and diagnosis (CADx). Such tools as convolutional neural networks (CNNs) can significantly decrease the surgeon's workload and make a better prognosis of patient conditions. CNNs need to be trained on a large amount of data in order to achieve a more trustworthy performance. In this study, transfer learning techniques, pre-trained CNNs, are adapted to a public dataset on osteosarcoma histological images to detect necrotic images from non-necrotic and healthy tissues. First, the dataset was preprocessed, and different classifications are applied. Then, Transfer learning models including VGG19 and Inception V3 are used and trained on Whole Slide Images (WSI) with no patches, to improve the accuracy of the outputs. Finally, the models are applied to different classification problems, including binary and multi-class classifiers. Experimental results show that the accuracy of the VGG19 has the highest, 96\%, performance amongst all binary classes and multiclass classification. Our fine-tuned model demonstrates state-of-the-art performance on detecting malignancy of Osteosarcoma based on histologic images. 

\end{abstract}


\end{frontmatter}

\input{01_Introduction.tex}
% %\input{02-NMGsStructure.tex}
\input{03_Methodology}

\input{04_Results.tex}

\input{05_Discussion.tex}

\bibliographystyle{elsarticle-num-names}
\bibliography{References.bib}






% \end{thebibliography}

\end{document}

%% file: 01_Introduction.tex
\section{Introducation}
\label{S:1.1}

Primary bone tumors account for 5-10\% of all new pediatric cancer diagnoses. Osteosarcoma is the most common form of malignant primary bone tumor.  Despite the limited approximately 1,000 new cases every year in the United States, the prognosis of osteosarcoma remains a challenging issue \cite{Chou2008}. There are two age peaks of incidence among patients, with a peak age of children under age 10, and adolescents at age 10-20 \cite{Arndt1999}. Osteosarcoma cancer usually occurs in the metaphysis of long bones on lower limbs, consisting of 40-50\% of the total cases \cite{Chou2008}. The symptoms of osteosarcoma usually begin with mild localized bone pain, redness, and warmth at the site of the tumor. Patients experience increasing pain, which often affects patients' movement and joint functions. If the early phase of osteosarcoma is not treated, it is expected to see a wide range of metastasis such as at lungs, other bones and soft tissues \cite{Lin}.

Histological biopsy tests, X-ray tests and magnetic resonance images are essential diagnosis to of osteosarcoma. Currently, the diagnosis of osteosarcoma includes a detailed history taking and physical examinations \cite{Wittig, Geller}. The presenting symptoms typically include deep-seated, constant, gnawing pain and swelling at the effected site. Pain in multiple areas may portend skeletal metastasis; therefore, they should be investigated appropriately \cite{Geller}.Beyond the examination, the standard studies for evaluation of potential osteosarcoma are laboratory tests, an X-ray of the entire affected bone, a magnetic resonance imaging (MRI) scan of the entire affected bone, a chest X-ray, a chest computed tomography (CT) scan, a whole-body technetium bone scan, and a percutaneous image-guided biopsy \cite{Geller}. Although the biopsy-based methods can effectively discover the malignancy, limitations in histological-guided biopsies and MRI scans have limited detecting capacity. Additionally, the preparation of histological specimens is time-consuming. For example, an accurate detection of osteosarcoma malignancy requires preparation of at least 50 histology slides to represent a plane of a large three-dimensional tumor \cite{Arndt1999}.

Due to the rise of cancer incidence and patient-specific treatment options, diagnosis and treatment of cancer are becoming more complex \cite{WANG20191686}. Pathologists must spend an extremely long time examining a large number of slides. Detecting the nuances of histological images can be difficult \cite{Picci}. Misdiagnosis often occurs due to the extensive work that decreases the accuracy of diagnosis. The osteoblasts' morphology has little difference in differentiated cells, which makes the image barely distinguishable. Also, the biopsy is a vital and time-consuming step to determine the presence of malignant tissue. Meanwhile, Computer-Aided Detection (CAD) technology offers a solution for radiologists to automatically detect malignancies \cite{Castellino2005}.

To address these limitations, microscopic image-based analysis has been the foundation of cancer diagnosis in recent years \cite{Geller}. However, it was not practical before the 2000s because of relatively low detection accuracy. The poor performance of CAD made clinical implementation impractical until the recent advances in computerized image detection \cite{MADABHUSHI}.

Recent advances enabled the possibility of turning histological slides to digital image datasets, in which machine learning can intervene on digital images to address some of the limitations. With the advent of whole slide imaging (WSI), digital pathology has become a part of the routine procedure in clinical diagnosis. The emergence of digital pathology provides new chances of developing new algorithms and software. A histological image can be quantified in such a system in order to improve the pathological procedures. The system digitizes glass slides with stained tissue sections at very high-resolution images, which makes computerized image analysis viable \cite{litjens2016deep}.

The primary goals of this study are: 

1) To demonstrate that the development of deep learning-based tools is capable of detecting the osteosarcoma malignancy with high accuracy based on a public dataset. The purpose is to successfully distinguish the typical patterns of non-tumors, necrotic tumors and viable tumors.

2) To explore a suitable deep learning framework for accurate detection and discover possible clues that contribute to performance.
To achieve the goals, histological medical image analysis based on transfer learning were applied to the pathology archives at Children’s Medical Center’s dataset \cite{dataset}. Two modified transfer learning approaches including VGG19 \cite{Simonyan2015VeryDC} and Inception V3 \cite{Szegedy_2015_CVPR} models were applied to the data. Compared to the previous results, we achieved an overall 2\% improvement in accuracy. The novelty of the model is that is being applied to different categories of the dataset and using the whole tile image as the input. 

\section{Literature Review}
\label{S:1.2}

Computer-aided technology in radiological and biopic detection becomes viable since 2010 \cite{Shin2016}. Remarkable progress has been achieved in medical images, primarily due to the availability of large-scale datasets and deep convolutional neural networks (CNNs) in the computer science area \cite{rawat2017deep}. This technology has been widely applied to a variety of medical images for the detection of different diseases, such as chest X-ray pneumonia, breast cancer, pulmonary edema, pulmonary fibrosis, gastric endoscopic images for celiac diseases and gastric cancer \cite{WANG20191686, serag2019translational}. The x-ray and biopsy for osteosarcoma share a similar pattern with these diseases; Therefore, it is practically feasible to use CNN to detect the early stage tissue morphological change. To decrease the mortality, it is imperative to prevent the early stage tumor from metastasis. Early automatic detection can not only decrease the chance of misdiagnosis but also serve as an assistant tool for the surgeon's preference to determine if metastasis has occurred. We believe the adoption of computer-aided technology using CNN can significantly reduce the surgeon's workload and achieve a better diagnosis of patients. 

Several state-of-the-art studies based on deep learning has been recognized as a recent major enhancement in histological image detection; However, most efforts of image detection are focused on histological images of breast cancer. In 2017, Jongwon \cite{Chang2017} did a pilot study on histopathology of breast cancer, which achieves an AUC value of 93\% on microscopic biopsy images in classifying benign or malignant tumors. They show that transfer learning is a viable and pre-trained model that is useful in classifying histological images. Erkan’s result \cite{Deniz2018} shows the state-of-the-art performance using VGG16 and AlexNet models, with an average of 90.96±1.59\% accuracy. This also indicates the suitability of these models for image classification tasks.

Other examples of image classification in recent years use similar methods: Jonathan De Matos \cite{de2019double} used double transfer learning to classify histopathologic images. Noorul Wahab \cite{Wahab2019} aimed at a more challenging task of segmentation and detection of mitotic nuclei. They used a similar hybrid CNN model and achieves 76\% AUC value. Other examples include the prediction of pathological invasiveness in lung adenocarcinoma \cite{Yanagawa2019}, Classification of Liver Cancer Histopathology Images \cite{Sun2020} , and Automated invasive ductal carcinoma detection \cite{Celik2020}. 

In Harish Babu Arunachalam’s study \cite{Arunachalam}, the article reports the first fully automated tool to assess viable and necrotic tumor in osteosarcoma using histological images and deep learning models. The goal is to label the diverse regions of tissue into a viable tumor, necrotic tumor, and non-tumor. They employed both machine learning and deep learning models. The ensemble learning model achieved an overall accuracy of 93.3\% with class-specific accuracies of 91.9\% for non-tumor, 95.3\% for viable tumor, and 92.7\% for necrotic tumor.

In machine learning and data mining algorithms, the main premise is that training and potential data should be in the same space and distribution. The problem arises when we have no access to enough training data in the specific research domain. Hence, we can obtain the basic parameters for training our deep learning model from pre-trained networks applied to larger data sets from other domains. In these situations, knowledge-transferring significantly improves learning outputs if done efficiently while minimizing expensive data labeling efforts \cite{Pan2010}.

%% file: 03_Methodology.tex
\section{Methodology}
\subsection{Dataset}
\begin{figure}
    \centering
    \includegraphics[width=\textwidth,height=4.5cm]{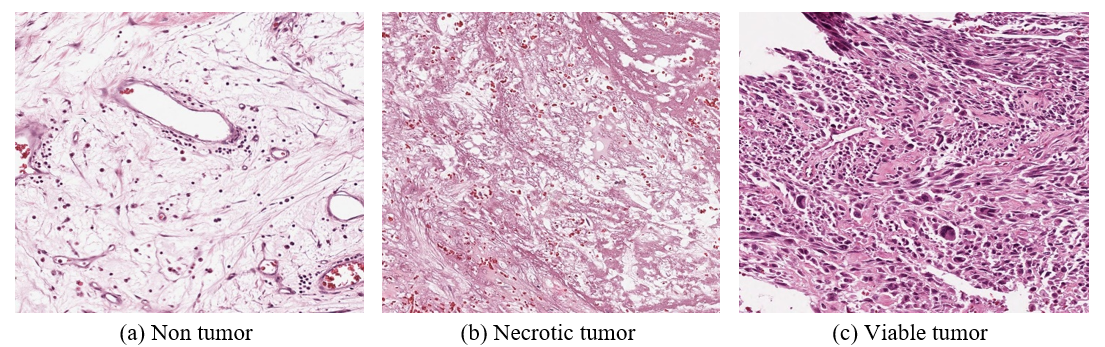}

    \caption{Sample Images from the dataset}\label{fig:sample_data}
\end{figure}

The dataset used in the study was obtained from the work of Arunachalam et al. where they provided a data set of osteosarcomas and conducted a variety of machine learning and deep learning techniques. Tumor samples from the Children's Medical Center, Dallas, were collected from the pathology reports of the osteosarcoma resection for 50 patients treated between 1995 and 2015. They selected 40 WSIs of the digitized images representing tumor heterogeneity and response properties in the study. In each WSI, 30 1024$\times$1024 pixel image tiles were randomly selected at the 10X magnification factor. 1,144 of the resulting 1,200 image tiles, such as those that fall into non-fabric, ink marks regions, and blurry images were chosen after removing irrelevant tiles. Moreover, they generated 56,929 patches of 128$\times$128 pixels. Some sample dataset images are shown in Figure \ref{fig:sample_data}.

\subsection{Data Preprocessing}
Original images of 1024$\times$1024 pixels were used for model training, validation, and evaluation. We split the datasets into training, validation, and testing images at a ratio of 70\%, 10\%, and 20\% respectively. The data are then augmented by using  image data generator of  “Keras”\cite{Keras}. In this step, all image intensities are first rescaled to the range of 0 to 1, and then the following augmentations have been performed: rotation, width shift, height shift, vertical flip, and horizontal flip. Due to memory limitations, we down sampled the original images by passing the input shape of 375$\times$375, rather than 1024$\times$1024.

\subsection{Model Selection}
There are 26 deep learning models in Keras Applications that can be used for prediction, feature extraction, and fine-tuning \cite{Keras}. Six of these models are applied for multi-class classification and among them we have chosen the best model for our experiment depending on the test accuracy. Table \ref{tab:Model_Selection} shows the test results of these models. VGG19 gives the best result among these models and we choose this model for future experiments.
\begin{table}[h]
    \centering
     \caption{Multi-class Result of Various Models}
    \begin{tabular}{|c|p{2cm}|p{2cm}|p{2cm}|p{2cm}|}
   
    \hline
     Model & Weighted Average Precision & Weighted Average Recall & Weighted Average F1-Score & Accuracy \\ 
    \hline 
  
    VGG16 &	0.89 & 0.88	& 0.88 & 0.883\\
    \hline
    VGG19 &	0.94 & 0.94 & 0.94 & 0.939 \\
    \hline
    ResNet50 & 0.22 & 0.47 & 0.30 &	0.470\\
    \hline
    InceptionV3 & 0.81 & 0.78 & 0.79 & 0.783\\
    \hline
    DenseNet201 & 0.61 & 0.58 & 0.56 & 0.583\\
    \hline
    NASNetLarge & 0.80 & 0.79 & 0.79 & 0.791\\
    \hline
 
    \end{tabular}
    \label{tab:Model_Selection}
\end{table}
\subsubsection{VGG19 Model}
\begin{figure}
    \centering
    \includegraphics[width=\textwidth,height=5.5cm]{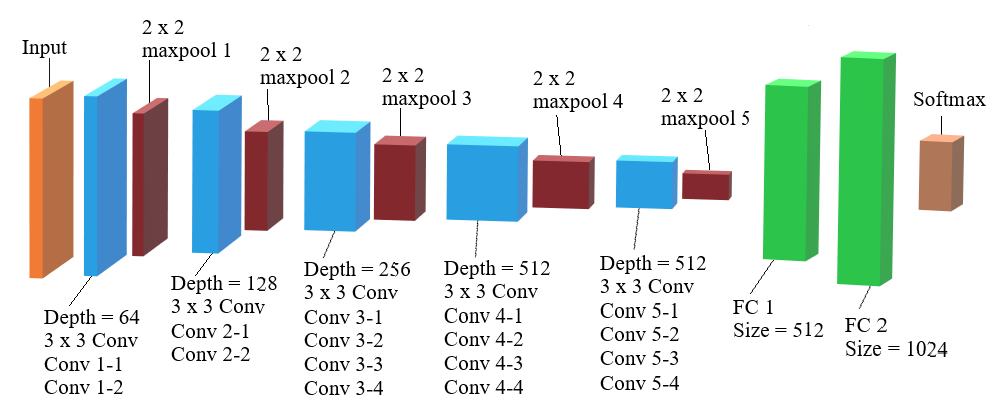}
    \caption{VGG19 Network Architecture }
    \label{fig:VGG19}
\end{figure}

We have used Keras applications for importing VGG19 model. Pre-trained weights have been used for model training. We have discarded the fully connected layer along with output layer of the VGG19 model. We have added two fully connected layers after the last “maxpool” layer. Dropout layers are used for avoiding over-fitting the training data. We have used “Relu” activation in the dense layers and “softmax” activation function in the output layer. Figure \ref{fig:VGG19} shows the VGG19 model architecture. All the “Conv 1-1” to “Conv 5-4”, and “maxpool 1” to “maxpool 5” use pre-trained weights. We have added the FC1, FC2, and softmax layers to this network. As shown in the figure, all the convolution layers use 3$\times$3 filters, and all the maxpooling layers use 2$\times$2 filters. The FC1 and FC2 layers contain 512 and 1024 neurons respectively. softmax layer’s neurons varies depending on our classification task. For binary and multi-class classification, it contains two and three neurons respectively.

%% file: 04_Results.tex
\section{Experimental Results}
\subsection{Setup}
With our dataset containing three classes, we performed four binary classifications and a multiclass (three classes) classification. In each classification, we applied two models: VGG19 and Inception V3. Inception V3 has been used for model comparison. The models are written in the Python programming language in the Keras deep learning framework. The models are trained and tested on a Nvidia GeForce RTX 2080Ti GPU platform. 

The loss functions used for binary classification and multiclass classifications are binary cross entropy and categorical cross entropy respectively. In both types of classification, Adam optimizer is applied for minimizing the loss function by updating the weight parameters. The learning rate is set to Keras’s default 0.01. Batch size is set to 80, 28, and 16 for training, validation, and testing respectively. All models are trained for 1500 epochs, with a callback that stops training when validation accuracy reaches over 98\%. 

Two-class classifications are evaluated on the following datasets: 1.) Non-Tumor (NT) versus Necrotic Tumor (NCT) and Viable Tumor (VT), 2.) Necrotic Tumor versus Non-Tumor, 3.) Viable Tumor versus Non-Tumor, and 4.) Necrotic Tumor versus Viable Tumor. We also performed the multiclass classification among the three classes: NT, NCT and VT. To evaluate our model performance, we presented confusion matrix, precision, recall, f1 score, and accuracy for all classifications. We also reported receiver operating characteristic (ROC) curve and area under the curve (AUC) for all the two-class classifications. 

Precision measures the percentage of correctly classified images in that specific predicted class, and recall measures the percentage of correctly classified images in the ground truth. F1 score is the weighted average of precision and recall. Accuracy measures the percentage of correctly classified (predicted) images among all the predictions. The receiver operating characteristic (ROC) curve shows the diagnostic ability of a binary classifier system for different thresholds. This curve plots the true positive rate (sensitivity) against false positive rate (1-specificity). The area under the curve (AUC) indicates that the classifier gives a randomly chosen positive instance higher probability than a randomly chosen negative instance.

\subsection{Results}
\begin{figure}
    \centering
    \includegraphics[width=\textwidth,height=12cm]{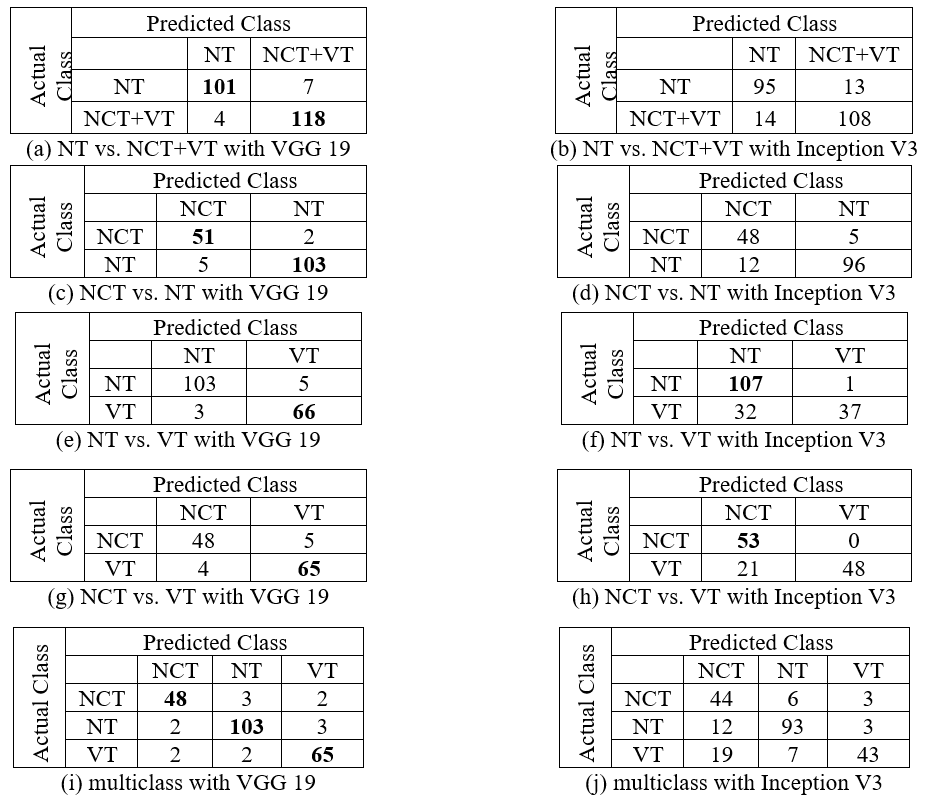}
    \caption{Confusion matrixes of all classifications. Here, NT = Non-Tumor, NCT = Necrotic Tumor, and VT = Viable Tumor}
    \label{fig:Confusion_Matrix}
\end{figure}

The evaluation metrics for all the classifications with two models are briefly presented in the following sections. Figure \ref{fig:Confusion_Matrix} shows the confusion matrix for all classifications with all three networks. 

\begin{figure}
    \centering
    \includegraphics[width=0.8\linewidth]{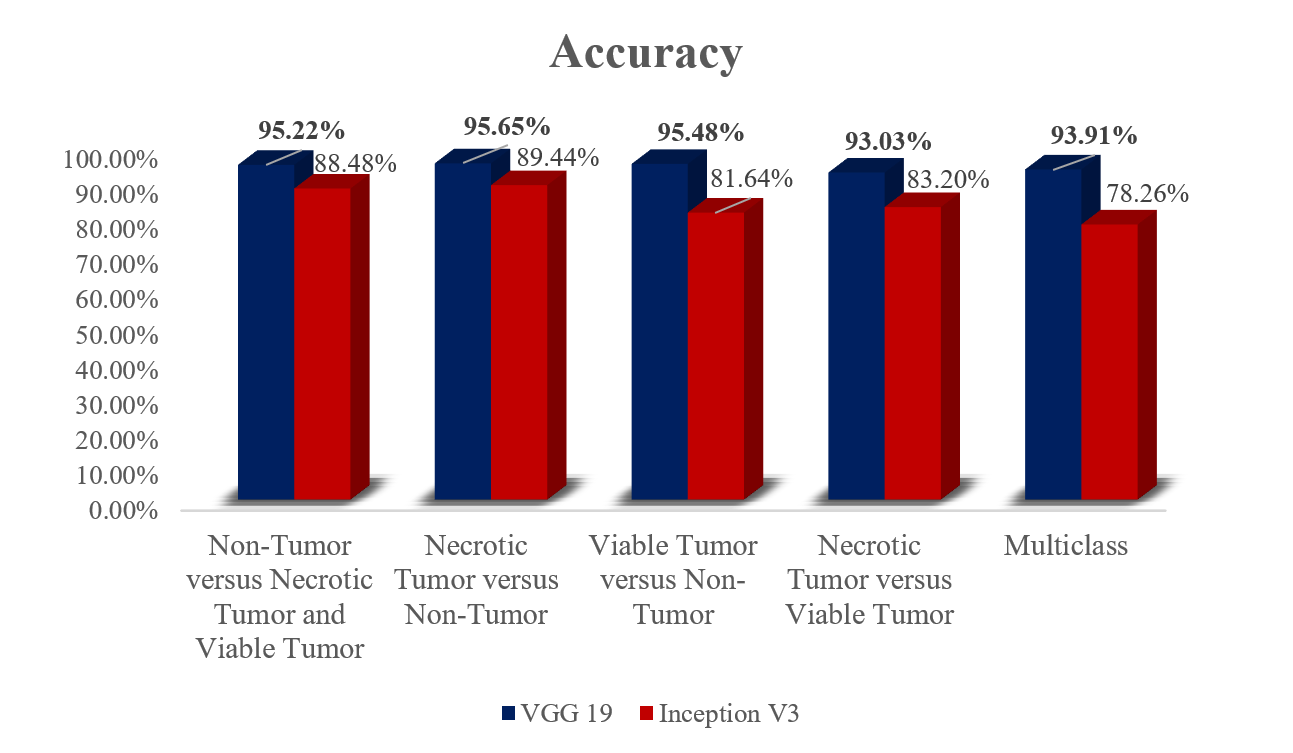}
    \caption{Accuracy Scores}
    \label{fig:accuracy}
\end{figure}

Table \ref{tab:binaryPrecision} and \ref{tab:multiclass_Precision} show the precision, recall, and f1 score for all the binary and multiclass classifications with each of the present networks. Figure \ref{fig:accuracy} shows the accuracy of the classifiers for all the classifications.
% Table binary precision and recall
\begin{table}[htbp]
  \centering
  \caption{Precision, Recall, and F1-Score for binary classes}
    \begin{tabular}{|c|c|c|c|c|c|c|c|}
     
    \hline
    \multicolumn{7}{|c|}{Non-Tumor versus Necrotic Tumor and Viable Tumor} \\
    \hline
     
    \multicolumn{1}{|c|}{} & \multicolumn{3}{c|}{Non-Tumor} & \multicolumn{3}{c|}{Necrotic and Viable Tumor} \\
    \hline
     
    \multicolumn{1}{|c|}{Networks} & \multicolumn{1}{c|}{Precision} & \multicolumn{1}{c|}{Recall} & \multicolumn{1}{c|}{F1} & \multicolumn{1}{c|}{Precision} & \multicolumn{1}{c|}{Recall} & \multicolumn{1}{c|}{F1} \\
    \hline
     
    \multicolumn{1}{|c|}{VGG19} & \multicolumn{1}{c|}{0.96} & \multicolumn{1}{c|}{0.94} & \multicolumn{1}{c|}{0.95} & \multicolumn{1}{c|}{0.94} & \multicolumn{1}{c|}{0.97} & \multicolumn{1}{c|}{0.96} \\
    \hline
     
    \multicolumn{1}{|c|}{Inception V3} & \multicolumn{1}{c|}{0.87} & \multicolumn{1}{c|}{0.88} & \multicolumn{1}{c|}{0.88} & \multicolumn{1}{c|}{0.89} & \multicolumn{1}{c|}{0.89} & \multicolumn{1}{c|}{0.89} \\
    \hline
     
    \multicolumn{7}{|c|}{Necrotic Tumor versus Non-Tumor} \\
    \hline
     
    \multicolumn{1}{|c|}{} & \multicolumn{3}{c|}{Necrotic Tumor} & \multicolumn{3}{c|}{Non-Tumor} \\
    \hline
     
    \multicolumn{1}{|c|}{Networks} & \multicolumn{1}{c|}{Precision} & \multicolumn{1}{c|}{Recall} & \multicolumn{1}{c|}{F1} & \multicolumn{1}{c|}{Precision} & \multicolumn{1}{c|}{Recall} & \multicolumn{1}{c|}{F1} \\
    \hline
     
    \multicolumn{1}{|c|}{VGG19} & \multicolumn{1}{c|}{0.91} & \multicolumn{1}{c|}{0.96} & \multicolumn{1}{c|}{0.94} & \multicolumn{1}{c|}{0.98} & \multicolumn{1}{c|}{0.95} & \multicolumn{1}{c|}{0.97} \\
    \hline
     
    \multicolumn{1}{|c|}{Inception V3} & \multicolumn{1}{c|}{0.8} & \multicolumn{1}{c|}{0.91} & \multicolumn{1}{c|}{0.85} & \multicolumn{1}{c|}{0.95} & \multicolumn{1}{c|}{0.89} & \multicolumn{1}{c|}{0.92} \\
    \hline
     
    \multicolumn{7}{|c|}{Viable Tumor versus Non-Tumor} \\
    \hline
     
    \multicolumn{1}{|c|}{} & \multicolumn{3}{c|}{Non-Tumor} & \multicolumn{3}{c|}{Viable Tumor} \\
    \hline
     
    \multicolumn{1}{|c|}{Networks} & \multicolumn{1}{c|}{Precision} & \multicolumn{1}{c|}{Recall} & \multicolumn{1}{c|}{F1} & \multicolumn{1}{c|}{Precision} & \multicolumn{1}{c|}{Recall} & \multicolumn{1}{c|}{F1} \\
    \hline
     
    \multicolumn{1}{|c|}{VGG19} & \multicolumn{1}{c|}{0.97} & \multicolumn{1}{c|}{0.95} & \multicolumn{1}{c|}{0.96} & \multicolumn{1}{c|}{0.93} & \multicolumn{1}{c|}{0.96} & \multicolumn{1}{c|}{0.94} \\
    \hline
     
    \multicolumn{1}{|c|}{Inception V3} & \multicolumn{1}{c|}{0.77} & \multicolumn{1}{c|}{0.99} & \multicolumn{1}{c|}{0.87} & \multicolumn{1}{c|}{0.97} & \multicolumn{1}{c|}{0.54} & \multicolumn{1}{c|}{0.69} \\
    \hline
     
    \multicolumn{7}{|c|}{Necrotic Tumor versus Viable Tumor} \\
    \hline
     
    \multicolumn{1}{|c|}{} & \multicolumn{3}{c|}{Necrotic Tumor} & \multicolumn{3}{c|}{Viable Tumor} \\
    \hline
     
    \multicolumn{1}{|c|}{Networks} & \multicolumn{1}{c|}{Precision} & \multicolumn{1}{c|}{Recall} & \multicolumn{1}{c|}{F1} & \multicolumn{1}{c|}{Precision} & \multicolumn{1}{c|}{Recall} & \multicolumn{1}{c|}{F1} \\
    \hline
     
    \multicolumn{1}{|c|}{VGG19} & \multicolumn{1}{c|}{0.92} & \multicolumn{1}{c|}{0.91} & \multicolumn{1}{c|}{0.91} & \multicolumn{1}{c|}{0.93} & \multicolumn{1}{c|}{0.94} & \multicolumn{1}{c|}{0.94} \\
    \hline
     
    \multicolumn{1}{|c|}{Inception V3} & \multicolumn{1}{c|}{0.72} & \multicolumn{1}{c|}{1} & \multicolumn{1}{c|}{0.83} & \multicolumn{1}{c|}{1} & \multicolumn{1}{c|}{0.7} & \multicolumn{1}{c|}{0.82} \\
    \hline
     
        %   &       &       &     &   &   &       &  \\
    \end{tabular}%
  \label{tab:binaryPrecision}%
\end{table}%

% multiclass precision and recall
\begin{table}[htbp]
  \centering
  \footnotesize
  \caption{Precision, Recall, and F1-Score for Multicalss }
    \begin{tabular}{|c|c|c|c|c|c|c|c|c|c|}
     
    \hline
    \multicolumn{10}{|c|}{Multiclass} \\
    \hline
     
          & \multicolumn{3}{c|}{Necrotic Tumor} & \multicolumn{3}{c|}{Non-Tumor} & \multicolumn{3}{c|}{Viable Tumor} \\
\hline
     
    Networks & Precision & Recall & F1    & Precision & Recall & F1    & Precision & Recall & F1 \\ \hline
     
    VGG19 & 0.92  & 0.91  & 0.91  & 0.95  & 0.95  & 0.95  & 0.93  & 0.94  & 0.94 \\
    \hline
     Inception V3 & 0.59  & 0.83  & 0.69  & 0.88  & 0.86  & 0.87  & 0.88  & 0.62  & 0.73 \\
    \hline
     
    \end{tabular}%
  \label{tab:multiclass_Precision}%
\end{table}%

\begin{figure}[ht]
\begin{subfigure}{.5\textwidth}
  \centering
  % include first image
  \includegraphics[width=.8\linewidth]{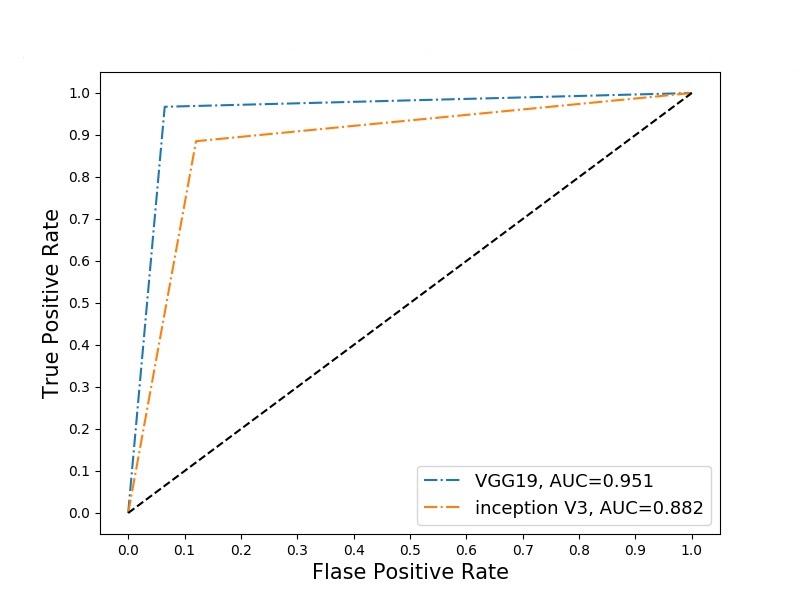}  
  \caption{Non-Tumor vs (Necrotic and Viable Tumor)}
  \label{fig:NT_Vs_NCT&VT}
\end{subfigure}
\begin{subfigure}{.5\textwidth}
  \centering
  % include second image
  \includegraphics[width=.8\linewidth]{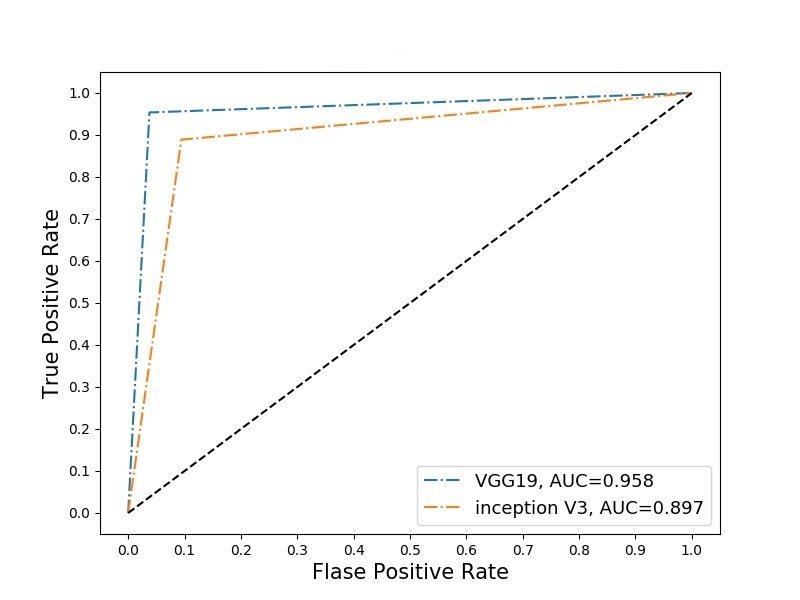}  
  \caption{Necrotic Tumor vs Non-Tumor}
  \label{fig:NCT_VS_NT}
\end{subfigure}
\begin{subfigure}{.5\textwidth}
  \centering
  % include third image
  \includegraphics[width=.8\linewidth]{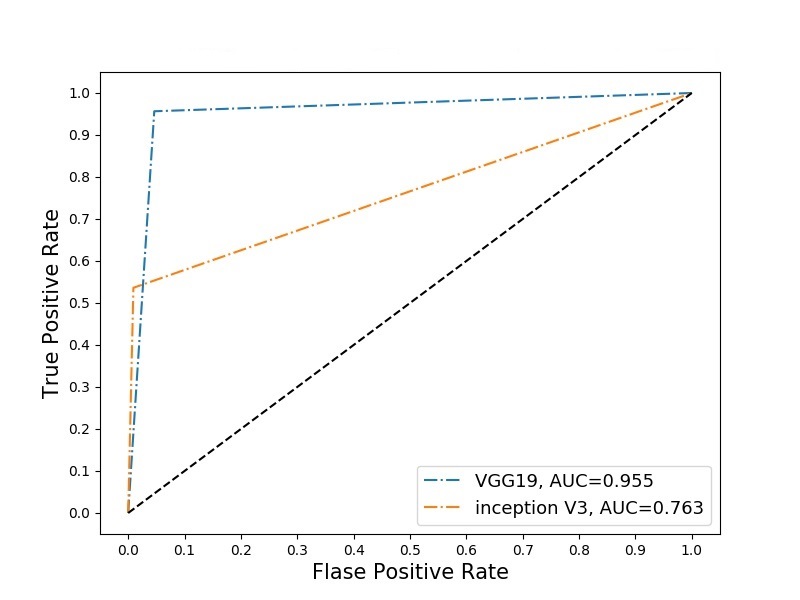}  
  \caption{Non-Tumor vs Viable Tumor}
  \label{fig:NT_VS_VT}
\end{subfigure}
\begin{subfigure}{.5\textwidth}
  \centering
  % include forth image
  \includegraphics[width=.8\linewidth]{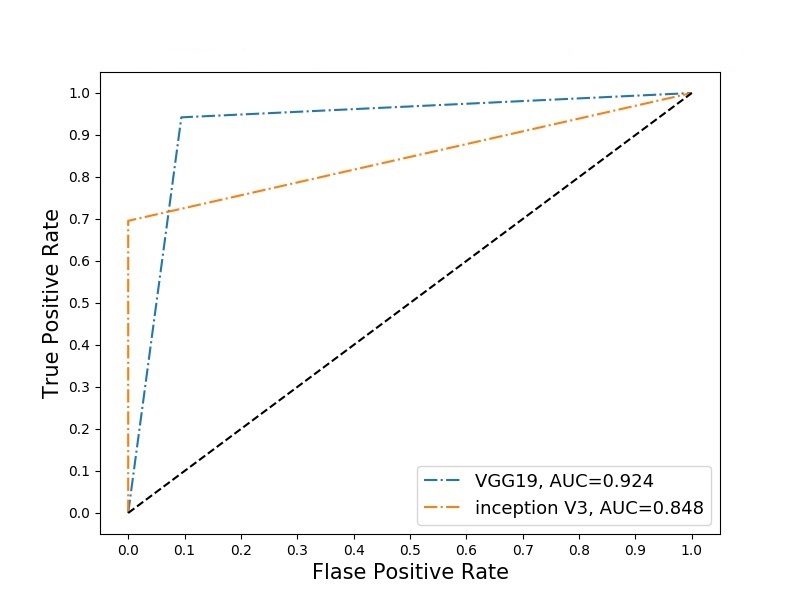}  
  \caption{Necrotic Tumor vs Viable Tumor}
  \label{fig:NCT_VS_VT}
\end{subfigure}
\caption{ROC and AUC of all two-class classifications}
\label{fig:roc_auc}
\end{figure}

%% file: 05_Discussion.tex
\section{Discussion}

Osteosarcoma is a common tumor in pediatric cases of cancer which requires extensive work of pathologists in order to confirm the case. While other medical images have already performed computerize analysis, osteosarcoma histological image is rarely mentioned in classification using deep learning models. We believe it is possible to make use of computer-aided technology to help classify and recognize the image of a malignant tumor. In this study, a deep learning-based technique has been used for image classification to detect the histologic images to identify malignancy of osteosarcoma. Our study provides an option of using a computer to accelerate the diagnosis and detection of osteosarcoma malignancy. Furthermore, we apply and compare two popular network architectures VGG19 and Inception V3\cite{Simonyan2015VeryDC, Szegedy_2015_CVPR}. Thus, we obtain higher performance than prior studies with the same dataset. We have configured and tested models with custom layers to achieve the best performance. 

From Figure \ref{fig:Confusion_Matrix}, we can see that for NT vs VT and NCT vs VT respectively the prediction of non-tumor and necrotic tumor is performed well by Inception V3. In all other cases VGG19 works very good compared to Inception V3. So, in overall balance VGG19 beats Inception V3. 

From Table \ref{tab:binaryPrecision} and \ref{tab:multiclass_Precision}, we can see that for VT vs NT and NCT vs VT cases precision of viable tumor and recall of necrotic tumor and non-tumor are high for Inception V3. But the interesting fact is that all the f1 scores are higher for VGG19 model. Since f1 score indicates the weighted average of precision and recall, a higher f1 score means precision and recall are close to each other for VGG19, where for inception V3 only a single metric is higher (either precision or recall) indicating lower score of the other one. Hence, in balance in overall performance, VGG19 beats inception V3 by a huge margin. From Figure \ref{fig:accuracy}, it is clear that for all classifications VGG19 achieves the highest accuracy. 

From Figure \ref{fig:roc_auc}, we can see that VGG19 has the highest AUC value for all binary (two-class) classifications. The AUC values are impressive (0.95, 0.96, 0.96, and 0.92 for non-tumor versus necrotic tumor and viable tumor, necrotic tumor versus non-tumor, viable tumor versus non-tumor, and necrotic tumor versus viable tumor classifications respectively), which assures us with great reliability. So, from all the above analytical discussion, it is safe to say that VGG19 works well for all classifications.
While Inception V3 has three types of convolutions (1$\times$1, 3$\times$3, 5$\times$5), VGG19 has only one type of convolution (3$\times$3). Instead of going deeper, Inception V3 goes wider on an image feature searching. As our dataset contains biopsy images in which some parts may only contain some specific features of a specific class (necrotic or viable), some of the inception kernels may not provide good features and in the concatenation layer, the performance may decrease. In VGG19, the kernel size is always the same (3$\times$3); which may lead to better classification accuracy specifically for our dataset. This dataset has a small number of images (1144), which is not suitable for deep learning models. Deep learning demands lots of data to learn the connection between given input and corresponding output. To overcome the data limitation problem, we applied transfer learning approach. Both VGG19 and inception V3 are pre-trained with the imagenet dataset, where all the low-level features (edge, curve etc.) are trained with imagenet dataset and we transfer that learned weights to our dataset. The fully connected layers and output layers are replaced in both models and trained with our dataset. 

To the best of our knowledge, this is the first pipeline that have been used in VGG19 and Inception architecture in Deep learning to recognize the osteosarcoma malignancy. The adjusted model can identify the minimal differences of images to predict the early signs of cancer. If the pipeline was deployed in various medical facilities, our model could help pathologists as an adjunct tool reducing their extensive work. 

The best accuracy is achieved by the VGG19 model compare to Arunachalam et al.’s deep learning model (a CNN model with three pairs of convolutions and pulling layers for sub-sampling, and two fully connected multi-layer perceptron). Table \ref{tab:comparison} represents the comparison of these two works. We have done a binary classification for all possible combinations between three classes, where \citet{Arunachalam}’s deep learning model provides a direct class specific accuracy. Therefore Table \ref{tab:comparison} represents our average accuracy for a specific tumor class. For viable tumor the average of  VT vs NT and NCT vs VT; for necrotic tumor the average of NCT vs NT and NCT vs VT; and for non-tumor the average of NT vs NCT and VT, NCT vs NT, and VT vs NT is represented. The comparison is done on the whole images (tile accuracy \cite{Arunachalam}), as we have used the 1144 whole images for our classification. Table \ref{tab:comparison} shows a better performance of non-tumor than other classes, which may be caused by the imbalance data in each class. This dataset contains 536, 345, and 263 whole images of non-tumor, viable tumor, and necrotic tumor respectively.

\begin{table}[htbp]
  \centering
  \caption{Result Comparison} 
    \begin{tabular}{|c|c|c|}  \hline
     
    \multirow{2}[4]{*}{Tumor type} & \multicolumn{2}{|c|}{Tile accuracy in \%} \\
    \cline{2-3}
         & VGG19 & \multicolumn{1}{c|}{Arunachalam\cite{Arunachalam}’s deep learning model} \\
     \hline
    Non-Tumor  & 95.45 & 89.5 \\ \hline
     
    Necrotic Tumor    & 94.34 & 91.5 \\ \hline
     
    Viable Tumor & 94.26 & 92.6 \\ \hline
     
    \end{tabular}%
  \label{tab:comparison}%
\end{table}%

Limitations include the lack of evaluation from pathologists. Even though our model reaches a high performance, it is suggested that the tool should be used under a pathologist's supervision. A further study is to compare our model’s performance with expert pathologists. The comparison can make sure this tool can detect new malignant cases in clinical practices. Besides, the existing data set might not indicate the future histological images from patients, therefore, the generalizability of our model might be problematic. To address this issue, it would be helpful to be adopted in medical facilities to assess its performance. 
\section{Conclusion}
Within the area of medical image processing, it is important to automate the classification of histological images by computer-aided systems. It is difficult and time-consuming to carry out a microscopic examination of histological images. Automatic diagnosis of histology alleviates the workload and enables pathologists to focus on critical cases. In this work, we used two pre-trained networks from Keras library, including VGG19 and InceptionV3. Regularization and optimization techniques were performed to avoid variance. The analyses were performed in two different ways, one binary classification, and the other one multi-class classification. VGG19 model achieved the highest accuracy in both binary and multi-class classifications, with an accuracy of 95.65\% and 93.91\% respectively. Furthermore, the highest F1 score in binary class belonged to the Necrotic Tumor versus Non-Tumor, 0.97. Our study compared to the previous study on the same data have outperformed both binary and multi-class. And finally, this study was the first usage of VGG19 and Inception V3 on the Osteosarcoma dataset, and the same framework can also be applied for other types of cancer.